\documentclass{iopart}
\usepackage{euscript,iopams,amssymb,amsfonts,graphicx,bm}
\usepackage{pgfplots,setstack}

\usepackage{float}
\usepackage{epsfig}
\usepackage{esint}
 \usepackage{tikz-cd}

\usepackage{bm,braket}

\eqnobysec

\newcommand{\R}{{\mathbb R}}
\newcommand{\Xo}{\overline{X}}
\newcommand{\XXo}{\overline{\mathbf X}}
\newcommand{\x}{\mathbf{x}}

\newcommand{\y}{\mathbf{y}}
\newcommand{\X}{\bm{X}}
\newcommand{\W}{\bm{W}}
\newcommand{\E}{\mathbb E}
\newcommand{\calU}{\mathcal{U}}
\newcommand{\calT}{\mathcal{T}}
\newcommand{\calA}{\mathcal{A}}

\newcommand{\calR}{\mathcal{R}}

\newcommand{\calM}{\mathcal{M}}

\newcommand{\z}{\zeta}
\renewcommand{\e}{{\rm e}}

\newcommand{\n}{\bm{n}}

\begin{document}
\title[Transition path theory and stochastic resetting]{Transition path theory for diffusive search with stochastic resetting} 

\author{Paul C. Bressloff}
\address{Department of Mathematics, Imperial College London, London SW7 2AZ, UK}

\begin{abstract}
Many chemical reactions can be formulated in terms of particle diffusion in a complex energy landscape. Transition path theory (TPT) is a theoretical framework for describing the direct (reaction) pathways from reactant to product states within this energy landscape, and calculating the effective reaction rate. It is now the standard method for analyzing rare events between long lived states. In this paper, we consider a completely different application of TPT, namely, a dual-aspect diffusive search process in which a particle alternates between collecting cargo from a source domain $A$ and then delivering it to a target domain $B$. The rate of resource accumulation at the target, $k_{AB}$, is determined by the statistics of direct (reactive or transport) paths from A to B. Rather than considering diffusion in a complex energy landscape, we focus on pure diffusion with stochastic resetting. Resetting introduces two non-trivial problems in the application of TPT. First, the process is not time-reversal invariant, which is reflected by the fact that there exists a unique non-equilibrium stationary state (NESS). Second, calculating $k_{AB}$ involves determining the total probability flux of direct transport paths across a dividing surface $S$ between $A$ and $B$. This requires taking into account discontinuous jumps across $S$ due to resetting. We derive a general expression for $k_{AB}$ and show that it is independent of the choice of dividing surface. Finally, using the example of diffusion in a finite interval, we show that there exists an optimal resetting rate at which $k_{AB}$ is maximized. We explore how this feature depends on model parameters.
\end{abstract}

    \maketitle

\newpage 
\section{Introduction}

\begin{figure}[b!]
\centering
\includegraphics[width=12cm]{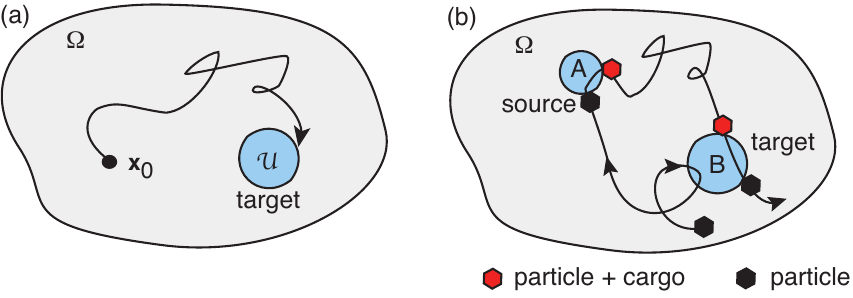}
\caption{(a) Classical search problem in which a diffusing particle searches for a target $\calU$ in a bounded domain $\Omega \subset \R^d$ with a totally reflecting exterior boundary $\partial \Omega$. The time to find the target is identified with the FPT to be absorbed at the target surface $\partial \calU$. After finding the target, the particle either loads or unloads a packet of resources. (b) Dual-aspect diffusive search process that alternates between the search for a resource domain A where a particle collects cargo and the subsequent search for a target domain B where the cargo is delivered. The rate of resource accumulation at the target is determined by the statistics of direct (reactive or transport) paths from A to B.}
\label{fig1}
\end{figure}

A classical problem in statistical physics is the diffusive search for some target ${\mathcal U}$ in a bounded domain $\Omega\subset \R^d$, see Fig. \ref{fig1}(a) \cite{Redner01,Benichou11,Bressloff13}. If the boundary $\partial \Omega$ is totally reflecting then the probability of eventually finding the target is unity. One typically formulates the search process as a first passage time (FPT) problem in which the target surface $\partial \calU$ is taken to be totally absorbing. The FPT is defined according to $\calT(\x_0)=\inf\{t >0, \X(t)\in \partial \calU|\X(0)=\x_0\}$, where $\X(t)$ is the position of the diffusing particle or searcher at time $t$ and $\x_0$ is its initial position. The mean FPT (MFPT) can be determined by solving a backward Kolmogorov equation or by calculating the probability flux into the target, whose Laplace transform is the generator of the FPT density. Higher-order moments can be calculated in a similar fashion. Various extensions include diffusive search within some energy landscape, diffusive search with stochastic resetting (as reviewed in Ref. \cite{Evans20}), and modifications in the absorption process itself. The last extension could involve taking the target surface $\partial \calU$ to be partially absorbing \cite{Grebenkov20}. Alternatively the whole domain $\calU$ could be partially absorbing, which means that the searcher freely enters and exits the target domain and can only be absorbed within the target interior \cite{Bressloff22a,Schumm21}.

There are two complementary interpretations of the search-and-capture process shown in Fig. \ref{fig1}(a): [I] the diffusing particle transports and delivers resources to the target (eg. intracellular vesicular transport) or [II] the particle searches for a target in order to extract resources from the target (eg. animal foraging). In this paper we consider a more complex diffusive search problem that is an alternating sequence of [I] and [II], see Fig.\ref{fig1}(b). The main idea is to assume that resources are initially located within a source domain A. This means that a particle first has to find the domain A in order to be supplied with a packet of resources (cargo). The resulting particle-cargo complex then searches for a target domain B in order to deliver its cargo and return to the bare particle state; if the complex returns to $A$ before reaching $B$ it does not load any additional cargo. In addition, we assume that the loading and unloading of cargo does not disrupt particle diffusion. (A more general model would allow for the particle to temporarily stop searching during the loading and unloading of cargo, which would introduce refractoriness into the model.) Finally, once the particle has delivered its cargo it continues searching for the source domain in order to load more cargo, and the process repeats (assuming that the amount of resources within $A$ and the capacity of the target $B$ are both unbounded)\footnote{This is of course an idealization in which the searcher is unable to memorize the locations of the source and target once they have been found for the first time. One possible scenario where a lack of memory might be advantageous is if there are multiple sources and multiple targets within the search domain, particularly given that a memory device costs resources.}.

One quantity of interest for the dual-aspect search process outlined in Fig. \ref{fig1}(b) is the mean rate at which the target accumulates resources. This can be calculated using the mathematical framework of transition path theory (TPT) \cite{Hummer04,Wienan06,Metzner06,Wienan11}. The latter was originally developed within the context of 
analyzing chemical reaction rates. In the latter case, the path traced out by a diffusing particle in Fig. \ref{fig1}(b) corresponds to the trajectory of a chemical system in some energy landscape, with $A$ representing reactant states and $B$ representing product states. One way to approximate the reaction rate is in terms of the mean crossing frequency of transitions from $A$ to $B$, which is  proportional to the total flux of reactive trajectories across any dividing surface $S$ separating the reactant states $A$ from the product states $B$. However, this clearly overestimates the reaction rate, since reactive trajectories can recross the surface $S$ many times during a single reaction, including multiple returns to the reactant states $A$ prior to reaching $B$. TPT deals with this overcounting by characterizing the statistical properties of the ensemble of reactive trajectories under the constraint that each reactive trajectory is a direct path from $A$ to $B$.

\begin{figure}[t!]
\centering
\includegraphics[width=10cm]{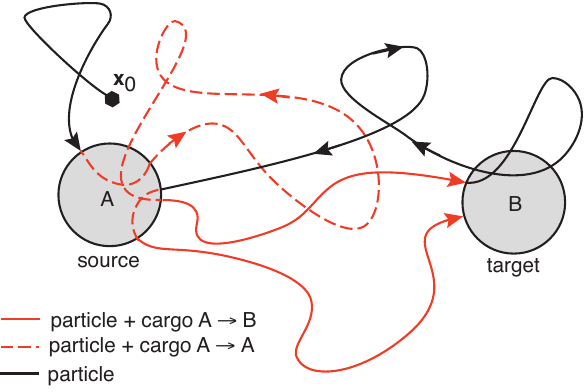}
\caption{Mapping of the sequential transport process to TPT. Direct paths or transport trajectories are shown as solid red curves. The effective rate $k_{AB}$ at which cargo is delivered from the source $A$ to the target B is determined by the number of transition trajectories $N_T$ observed over a time interval $[-T,T]$ according to $k_{AB}=\lim_{T\rightarrow \infty} N_T/2T$. (See text for further details.)}
\label{fig2}
\end{figure}

The mapping of the dual-aspect diffusive search problem to TPT is shown in Fig. \ref{fig2}. Let $\X(t)$, $-\infty<t <\infty$, denote the position of the searcher at time $t$. First, we distinguish between sections of the trajectory along which the particle is carrying cargo (red curves) from sections along which the particle is free (black curves). Thus any trajectory exiting $A$ is red, whereas any trajectory exiting $B$ is black. We further subdivide each red section into the part that forms a direct connection from $A$ to $B$ (solid red curves), which we will call a {\em transport trajectory}, and the remainder that returns to $A$ without reaching $B$ (dashed red curves). Let $N_T$ denote the number of transport trajectories observed during the time interval $[-T,T]$, then the mean frequency at which these trajectories are observed within the given trajectory is
\begin{equation}
k_{AB}=\lim_{T\rightarrow \infty} \frac{N_T}{2T}.
\end{equation}
Assuming that the stochastic process is ergodic, we can equate $k_{AB}$ with an ensemble average that is independent of the particular trajectory and can thus be calculated using TPT. Within the context of stochastic transport processes, $k_{AB}$ is equivalent to the mean frequency at which units of cargo are delivered to $B$. 

The mathematical analysis of TPT is typically developed within the context of stochastic differential equations (SDEs) as exemplified by a multi-dimensional Langevin equation \cite{Wienan06,Metzner06,Wienan11}. One of the novel aspects of stochastic transport processes is that the searcher dynamics may be described by a more general stochastic process \cite{Bressloff22}. Examples include active motor transport \cite{Newby10}, active Brownian and run-and-tumble particles \cite{Solon15},  facilitated diffusion \cite{Berg81,Kolomeisky11}, and L\`evy flights \cite{Bartumeus02}. In this paper we construct TPT for yet another example, namely, diffusion with stochastic resetting. That is, the position of the  particle resets to a fixed location $\x_r\in \Omega\backslash (A\cup B)$ at a constant rate $r$ \cite{Evans11a,Evans11b,Evans14}. It is well known that such a process is not time-reversal invariant, which is reflected by the fact that there exists a unique non-equilibrium stationary state (NESS). Such a state exists even if the dynamics is unbounded, that is, $\Omega =\R^d$. The application of TPT is thus non-trivial.

The structure of the paper is as follows. In section 2, we review TPT for an overdamped Brownian particle in $\R^d$ \cite{Wienan06,Metzner06,Wienan11}. The main analysis is developed in section 3, where we extend TPT to include the effects of stochastic resetting. First, we construct the reverse-time diffusion process with resetting by extending the formalism of Ref. \cite{Anderson82}. (This is necessary in order to determine the so-called backward committor function.) We then calculate the accumulation rate $k_{AB}$ by generalizing the derivations presented in the appendices of Ref. \cite{Metzner06}. In particular, we equate $k_{AB}$ with the time-averaged probability flux of transport trajectories across a dividing surface between the source and target domains. There are two distinct contributions to $k_{AB}$ corresponding, respectively, to paths that cross $S$ smoothly and those that jump across $S$ due to stochastic resetting. We also prove that $k_{AB}$ is independent of the choice of dividing surface $S$, and that it is a positive quantity. Finally, in section 4 we illustrate the theory by considering pure diffusion with resetting in a finite interval. We show that there exists an optimal resetting rate at which $k_{AB}$ is maximized, and explore how this feature depends on model parameters.

\section{TPT for overdamped Brownian motion}

Let $\X(t)\in \R^d$, $0\leq t \leq T$, denote the position of an overdamped Brownian particle evolving according to the SDE
\begin{equation}
dX_j(t)=f_j(\X(t))dt+\sqrt{2D}dW_j(t),
\label{SDE}
\end{equation}
where $D$ is the diffusivity and $W_j(t)$, $j=1,\ldots,d$, are independent Wiener processes. Let
\[p_0(\x,t)d\x={\mathbb P}[\x\leq \X(t)\leq \x+d\x],\]
with the initial condition $p_0(\x,0)=\overline{p}_0(\x)$. (The subscript on $p_0$ indicates that there is no stochastic resetting.) The probability density $p_0(\x,t)$ evolves according to the forward Fokker-Planck (FP) equation
\begin{eqnarray}
\label{fw}
 \frac{\partial p_0(\x,t)}{\partial t}=- \sum_{j=1}^d\frac{\partial f_j(\x)p_0(\x,t)}{\partial x_j}+D\sum_{j=1}^d\frac{\partial^2p_0(\x,t)}{\partial x_j^2}.
\end{eqnarray}
We will assume that for the given drift functions $f_j(\x)$, there exists a stationary measure $\rho_0(\x)d\x$ such that
\begin{equation}
\lim_{t\rightarrow \infty} p_0(\x,t)=\rho_0(\x).
\end{equation}
In particular, suppose that we have a conservative force for which
$ f_j(\x)=-\gamma^{-1}\partial_jU(\x)$, where $U(\x)$ is a potential energy function and $\gamma$ is a friction coefficient satisfying the Einstein relation $D\gamma = k_BT$. The stationary density is then given by the Boltzmann distribution
 $\rho_0(\x)=Z^{-1}\e^{-U(\x)/k_BT}$ with $Z=\int_{\R^d}\e^{-U(\y)/k_BT}d\y<\infty$. In addition, the diffusion process is time reversible and all equilibrium probability currents vanish.

\subsection{Ensemble of transport trajectories}

Let $\X(t)$, $-\infty < t <\infty$, be an infinitely long sample trajectory of the SDE (\ref{SDE}), which is ergodic with respect to the equilibrium probability density $\rho_0(\x)$. That is, given any suitable observable $\Phi(\x)$, we have the equivalence of time and ensemble averages:
\begin{equation}
\lim_{T\rightarrow \infty} \frac{1}{2T}\int_{-T}^{T} \Phi(\X(t))dt =\int_{\R^d}\Phi(\x)\rho_0(\x)d\x.
\label{ergodic}
\end{equation}
Suppose that $A\subset \R^d$ and $B\subset \R^d$ are two bounded regions in the phase space $\R^d$ that specify the source and target domains, respectively, see Fig. \ref{fig2}. Any trajectory $\X(t)$, $-\infty< t <\infty$, can be partitioned into pieces that are either transport trajectories or their complement. Each transport trajectory connects $\partial A$ to $\partial B$. That is, it starts at a point on the boundary $\partial A$ and ends at a point on the boundary $\partial B$ without ever returning to $A$. It follows from ergodicity that the set of all transport trajectories forms an ensemble $\Gamma$ whose statistical properties are independent of the particular trajectory $\X(t)$ used to generate the ensemble. Suppose that $\X(t)\notin A\cup B$ at time $t$ and set
\begin{eqnarray}
T^-_{AB}(t)&=\sup\{ t'\leq t \mbox{ such that } \X(t') \in A\cup B\},\\
T^+_{AB}(t)&=\inf\{ t'\geq t \mbox{ such that } \X(t') \in A\cup B\}.
\end{eqnarray}
Then
\begin{eqnarray}
\Gamma=\{\X(t); \, \X(t)\notin A\cup B,\, \X(T^-_{AB}(t))\in A,\,\X(T^+_{AB}(t))\in B\} .
\end{eqnarray}

One of the major objects of interest in transition path theory is the probability density that a trajectory passing through $\x \notin A\cup B$ at time $t$ is a transport trajectory at time $t$. For a given trajectory $\X(t)$, let ${\mathcal R}$ denote the set of times for which $\X(t)$ is on a transport trajectory. The probability density of transport trajectories $\rho_{AB}^{(0)}(\x)$ is defined according to
\begin{equation}
\lim_{T\rightarrow \infty} \frac{\int_{{\mathcal R}\cap [-T,T]} \Phi(\X(t))dt}{\int_{{\mathcal R}\cap [-T,T]}dt}=\int_{\Omega_{AB}}\Phi(\x)\rho_{AB}^{(0)}(\x)d\x,
\end{equation}
where $\Omega_{AB}=\R^d\backslash(A\cup B)=(A\cup B)^c$. It can be proven that \cite{Wienan06,Metzner06,Wienan11}
\begin{equation}
\fl \rho_{AB}^{(0)}(\x)=\frac{1}{Z_{AB}^{(0)}}q_0(\x)\overline{q}_0(\x) \rho_0(\x),\quad Z_{AB}^{(0)}=\int_{\Omega_{AB}} q_0(\x)\overline{q}_0(\x) \rho_0(\x)d\x,
\end{equation}
where $q_0(\x)$ is the probability that the transport trajectory reaches first $B$ before $A$, and $\overline{q}_0(\x)$ is the probability that the transport trajectory came from $A$ rather than $B$.

\subsection{Committor functions} The probabilities $q_0(\x)$ and $\overline{q}_0(\x)$ are known as the forward and backward committor functions. The former is equivalent to the splitting probability that a trajectory starting at $\x$ reaches $B$ before $A$ and thus satisfies the backward Kolmogorov equation
\numparts
\begin{eqnarray}
\label{qa}
 &\sum_{j=1}^d f_j(\x)\frac{\partial q_0(\x)}{\partial x_j}+D\sum_{j=1}^d\frac{\partial^2q_0(\x)}{\partial x_j^2}=0,\\ &  q_0(\x)=0 \mbox{ for } \x \in \partial A,\  q_0(\x)=1 \mbox{ for } \x \in \partial B.
 \label{qb}
\end{eqnarray}
\endnumparts
Determining the corresponding Kolmogorov equation for the backward committor function is more involved. Intuitively speaking, we can identify $\overline{q}_0(\x)$ as the splitting probability of the corresponding reverse-time diffusion process that starts at $\x$ and reaches $A$ before $B$.
Hence, the nontrivial step is determining the evolution equation for the reverse-time diffusion process.  Let $\XXo(t)=\X(T-t)$. In the absence of resetting and under mild conditions on the drift vector ${\bf f}(\x)$ and initial density $p_0(\x)$, it can be proven that the reverse-time process $\XXo(t)$ satisfies an SDE of the form \cite{Anderson82} (see also section 3.2)
\begin{equation}
d\Xo_j(t)=\overline{f}_j(\XXo(t),t)dt+\sqrt{2D}dW_j(t),
\end{equation}
where
\begin{equation}
\overline{f}_j(\x,t)=-f_j(\x)+\frac{2D}{p_0(\x,T-t)} \frac{\partial}{\partial x_j}p_0(\x,T-t).
\label{barft}
\end{equation}
In particular, taking $p_0(\x)=\rho(\x)$, we have
\begin{equation}
d\XXo_j(t)=\overline{f}(\XXo(t))dt+\sqrt{2D}dW_j(t),
\end{equation}
with
\begin{equation}
\overline{f}_j(\x)=-f_j(\x)+\frac{2D\partial_{x_j}\rho_0(\x)}{\rho_0(\x)}.
\label{barf}
\end{equation}
The corresponding forward FP equation for the reverse-time process is
\begin{eqnarray}
\label{fwrev}
 \frac{\partial \overline{p}(\x,t)}{\partial t}=- \sum_{j=1}^d\frac{\partial \overline{f}_j(\x)\overline{p}_0(\x,t)}{\partial x_j}+D\sum_{j=1}^d\frac{\partial^2\overline{p}_0(\x,t)}{\partial x_j^2},
\end{eqnarray}
where
\[\overline{p}_0(\x,t)d\x={\mathbb P}[\x\leq \XXo(t)\leq \x+d\x].\]
It now follows that $\overline{q}(\x)$ satisfies the backward Kolmogorov equation
\numparts
\begin{eqnarray}
\label{qbara}
& \sum_{j=1}^d\overline{f}_j(\x)\frac{\partial \overline{q}_0(\x)}{\partial x_j}+D\sum_{j=1}^d\frac{\partial^2\overline{q}_0(\x)}{\partial x_j^2}=0,\\ & \overline{q}_0(\x)=1 \mbox{ for } \x \in \partial A,\quad \overline{q}_0(\x)=0 \mbox{ for } \x \in \partial B.
\label{qbarb}
\end{eqnarray}
\endnumparts
In the particular case of an overdamped Brownian particle subject to a conservative force, we have a reversible diffusion process for which $\overline{q}_0(\x)=1-q_0(\x)$. This follows from substituting for $\rho_0(\x)$ in (\ref{barf}) using the Boltzmann distribution:
\begin{eqnarray}
\fl \overline{f}_j(\x)=-f_j(\x)+\frac{2D\partial_{x_j}\e^{-U(\x)/k_BT}}{e^{-U(\x)/k_BT}}=-f_j(\x)-\frac{2D}{k_BT} \partial_{x_j}U(\x) =f_j(\x).
\end{eqnarray}
However, if the force is non-conservative force or stochastic resetting is included (see section 3.1), then time reversibility no longer holds. Finally, note that although the boundary value problems for the committor functions are defined in $\R^d\backslash A\cup B$, we extend their domains of definition by taking
\begin{equation}
\fl q_0(\x)=0,\bar{q}_0(\x)=1 \mbox{ for all } \x \in A,\quad q_0(\x)=1,\bar{q}_0(\x)=0 \mbox{ for all } \x \in B.
\end{equation}

\subsection{Probability current and transition rate} Another quantity of interest is the probability current ${\bf J}_{AB}^{(0)}(\x)$ of transport trajectories crossing a dividing surface $ S\subset \Omega_{AB}$, with $A$ on one side and $B$ on the other side of the surface. (The superscript $(0)$ again indicates that there is no resetting.) Integrating ${\bf J}_{AB}^{(0)}(\x) $ over $S$ yields the total probability flux of transport trajectories across this surface, which determines the target accumulation rate $k_{AB}^{(0)}$ according to
\begin{equation}
k_{AB}^{(0)}=\int_S{\bf J}_{AB}^{(0)}(\x) \cdot \n(\x)d\sigma(\x).
\label{KAB}
\end{equation}
The vector $\n(\x)$ denotes the unit normal to $S$ pointing towards $B$ and $d\sigma(\x)$ is the surface element on $S$. It can be proven that \cite{Metzner06} (see section 3.2)
\begin{eqnarray}
\fl J_{AB,j}^{(0)}(\x) =q_0(\x)\overline{q}_0(\x) J_j^{(0)}(\x)+D\overline{q}_0(\x)\rho_0(\x) \frac{\partial q_0(\x)}{\partial x_j}-D q_0(\x)\rho_0(\x) \frac{\partial \overline{q}_0(\x)}{\partial x_j},
\label{JAB}
\end{eqnarray}
where ${\bf J}^{(0)}(\x)$ is the equilibrium probability current with components
\begin{equation}
J_j^{(0)}(\x)=f_j(\x)\rho_0(\x) -D \frac{\partial \rho_0(\x)}{\partial x_j}.
\label{Jj}
\end{equation}
Note that ${\bf J}_{AB}^{(0)}(\x)=0$ for all $\x \in A\cup B$.
Moreover, the reaction rate can be rewritten as the volume integral
\begin{equation}
k_{AB}^{(0)}=D\int_{\Omega_{AB}} \rho_0(\x)\sum_{j=1}^d\left (\frac{\partial q_0(\x)}{\partial x_j}\right )^2 d\x.
\label{KAB2}
\end{equation}
In the particular case of a conservative force, we have $J_j^{(0)}(\x)=0$ for all $j=1,\ldots,d$ and $\overline{q}_0(\x)=1-q_0(\x)$, which means that 
\begin{eqnarray}
J_{AB,j}^{(0)}(\x)=DZ^{-1}\e^{-U(\x)/k_BT} \frac{\partial q_0(\x)}{\partial x_j}.
\end{eqnarray}

\section{Overdamped Brownian motion with stochastic resetting}

Now suppose that the Brownian particle resets to a fixed position $\x_r$ at a rate $r$, see Fig. \ref{fig3}. 
When resetting is included, the SDE is modified according to
\begin{equation}
\fl d\X(t)=\left \{ \begin{array}{ll} \x_r -\X(t)& \mbox{ with probability } rdt\\
f(\X(t))dt+\sqrt{2D}d\W(t) & \mbox{ with probability } (1-rdt)\end{array}\right . .
\label{SDEr}
\end{equation}
The probability density $p(\x,t)$ evolves according to the modified forward FP equation \cite{Evans11a,Evans11b,Evans14}
\begin{eqnarray}
\label{fwr}
\fl  \frac{\partial p(\x,t)}{\partial t}=- \sum_{j=1}^d\frac{\partial f_j(\x)p(\x,t)}{\partial x_j}+D\sum_{j=1}^d\frac{\partial^2p(\x,t)}{\partial x_j^2}-rp(\x,t)+r \delta(\x-\x_r),\end{eqnarray}
with $ p(x,0) = \overline{p}(x)$. Similarly, we define the propagator $p(\x,t|\y,t_0)$ as the solution to equation (\ref{fwr}) under the initial condition $p(\x,t_0|\y,t_0)=\delta(\x-\y)$. The propagator also satisfies the backward FP equation
 \begin{eqnarray}
  \fl \frac{\partial p(\x,t|\y,t_0)}{\partial t}&= -\frac{\partial p(\x,t|\y,t_0)}{\partial t_0}\nonumber\\
 \fl  &=\sum_{j=1}^d f_j(\y) \frac{\partial p(\x,t|\y,t_0)}{\partial y_j}+D\sum_{j=1}^d\frac{\partial^2p(\x,t|\y,t_0)}{\partial y_j^2} \nonumber \\
 \fl  & \qquad -rp(\x,t|\y,t_0)+r p(\x,t|\x_r,t_0).
  \label{bwr}
\end{eqnarray}
(Under time translation invariance, we have $p(\x,t|\y,t_0)=p(\x,t-t_0|\y,0)$.) 

In the absence of stochastic resetting, ergodicity with respect to a stationary density $\rho_0(\x)$ depends on the properties of the force vector ${\bf f}(\x)$. If the external force field is zero (flat energy landscape), then Brownian motion in an unbounded domain is non-ergodic, reflecting the fact that the stationary density is zero pointwise. One of the important consequences of stochastic resetting is the existence of a nontrivial stationary density $\rho(\x)=\lim_{t\rightarrow \infty} p(\x,t|\x_0,0)$ for unbounded Brownian motion \cite{Evans11a,Evans11b,Evans14}. In particular, $\rho(\x)$ represents an NESS because there exist non-zero probability fluxes. That is the point $\x_r$ acts as a probability source, whereas all positions $\x\neq \x_r$ are potential probability sinks. An immediate issue is whether or not the resulting stochastic process is ergodic with respect to the NESS. The ergodicity of diffusion processes with stochastic resetting has recently been explored in  a number of studies \cite{Stojkoski21,Wang21,Stojkoski22,Wang22,Barkai23}. Although it has not been proven rigorously, normal diffusion processes with Poissonian resetting appear to be ergodic, and we will assume this in the following.

Taking the large-$t$ limit of equation (\ref{fwr}) shows that $\rho(\x)$ (if it exists for a given force field ${\bf f}(\x)$) satisfies the stationary equation
\begin{eqnarray}
\label{fwrstat}
\fl  0 =- \sum_{j=1}^d\frac{\partial J_j(\x)}{\partial x_j}-r\rho(\x)+r \delta(\x-\x_r),\quad J_j(\x)=f_j(\x)\rho(\x)-D\frac{\partial \rho(\x)}{\partial x_j}.\end{eqnarray}
It immediately follows that the current ${\bf J}(\x)$ is not divergence-free.
An alternative way to determine the NESS is to note that the propagator satisfies the last renewal equation \cite{Evans20}
\begin{equation}
\label{eq7:renewal}
p(\x,t|\x_0,0)=\e^{-rt}p_0(\x,t|\x_0,0)+r\int_{0}^tp_0(\x,\tau|\x_r,0)\e^{-r\tau}d\tau,
\end{equation}
where $p_0$ is the corresponding propagator without resetting.
The stationary state $\rho(\x)$ is obtained by taking the limit $t\rightarrow \infty$ in equation (\ref{eq7:renewal}):
\begin{equation}
\label{sas}
\rho(\x) =r\int_0^{\infty}p_0(\x,\tau|\x_r,0)\e^{-r\tau} d\tau.
\end{equation}
That is, $\rho(\x)$ is determined by the $r$-Laplace transform of the propagator $p_0$ (assuming it exists). In addition, the backward FP equation (\ref{bwr}) implies that the forward committor function $q(\x)$ satisfies the boundary value problem
\begin{equation}
{\mathbb L}_rq(\x)=0,\quad q(\x)=0 \mbox{ for } \x \in \partial A,\quad q(\x)=1 \mbox{ for } \x \in \partial B,
\label{rf}
\end{equation}
with
\begin{eqnarray}
{\mathbb L}_rq(\x)\equiv \sum_{j=1}^d f_j(\x) \frac{\partial q(\x)}{\partial x_j}+D\sum_{j=1}^d\frac{\partial^2q(\x)}{\partial x_j^2}  -rq(\x)+r q(\x_r).
\label{rf2}
 \end{eqnarray}
In order to determine the corresponding Kolmogorov equation for the backward committor function, we need to derive the FP equation for the reverse-time diffusion process with resetting.

\subsection{Reverse-time diffusion process with resetting and the backward committor function}

We obtain the reverse-time Markov process by extending the approach presented in Ref. \cite{Anderson82}. Consider the joint probability density
\begin{equation}
\rho(\x,t,\y,t_0)=p(\x,t|\y,t_0)p(\y,t_0),\quad t_0<t,
\end{equation}
where $p(\y,t_0)$ evolves according to equation (\ref{fwr}) with $\x\rightarrow \y$,  $t\rightarrow t_0$ and $p(\y,0)=p_0(\y)$. Differentiating both sides with respect to $t_0$ we have
\begin{eqnarray}
\fl\frac{\partial \rho}{\partial t_0}&=\frac{\partial p(\x,t|\y,t_0)}{\partial t_0}p(\y,t_0)+\frac{\partial p(\y,t_0)}{\partial t_0}p(\x,t|\y,t_0)\nonumber \\
\fl&=-\bigg [\sum_{j=1}^df_j(\y)\frac{\partial p(\x,t|\y,t_0)}{\partial  y_j}+D\sum_{j=1}^d\frac{\partial^2p(\x,t|\y,t_0)}{\partial y_j^2}\nonumber \\
\fl &\qquad -rp(\x,t|\y,t_0)+r p(\x,t|\x_r,t_0)\bigg] p(\y,t_0)\nonumber \\
\fl&-\left [\sum_{j=1}^d \frac{\partial f_j(\y)p(\y,t_0)}{\partial y_j}-D\sum_{j=1}^d\frac{\partial^2p(\y,t_0)}{\partial y_j^2} +rp(\y,t_0)-r \delta(\y-\x_r)\right ]p(\x,t|\y,t_0)\nonumber \\
\fl&=-p(\y,t_0)\sum_{j=1}^df_j(\y)\frac{\partial[\rho(\x,t,\y,t_0)/p(\y,t_0)]}{\partial y_j}\nonumber \\
\fl &\quad -Dp(\y,t_0)\sum_{j=1}^d\frac{\partial^2[\rho(\x,t,\y,t_0)/p(\y,t_0)]}{\partial y_j^2} +r[\delta(\y-\x_r)-p(\y,t_0)]p(\x,t|\x_r,t_0) \nonumber \\
\fl&\quad +\frac{1}{p(\y,t_0)}\left [-\sum_{j=1}^d \frac{\partial f_j(\y)p(\y,t_0)}{\partial y_j}+D\sum_{j=1}^d\frac{\partial^2p(\y,t_0)}{\partial y_j^2}   \right ]\rho(\x,t,\y,t_0)\nonumber \\
\fl &=- \sum_{j=1}^d\frac{\partial f_j(\y)\rho(\x,t,\y,t_0)}{\partial y_j}-D\sum_{j=1}^d \frac{\partial^2\rho(\x,t,\y,t_0)}{\partial y_j^2} \nonumber \\\fl &\quad +2D\sum_{j=1}^N\frac{\partial}{\partial y_j}\left [\frac{1}{p(\y,t_0)}\frac{\partial p(\y,t_0)}{\partial y_j}\rho(\x,t,\y,t_0)\right ]\nonumber \\
\fl &\quad -\left [\omega(\y|\x_r,t_0)-\delta(\y-\x_r)\int_{\infty}^{\infty} \omega(\z|\x_r,t_0)d\z\right ]\rho(\x,t,\x_r,t_0).
\label{re}
\end{eqnarray}
The last line represents jumps $\x_r \rightarrow \y$ with transition rate
\begin{equation}
\omega(\y|\x_r,t_0)=\frac{rp(\y,t_0)}{p(\x_r,t_0)}.
\end{equation}
We can reinterpret equation (\ref{re}) as the forward FP equation for a time reversed process by setting
$\overline{p}(\x,t|\X,0)=\rho(\X,T,\x,T-t)/p(\X,T)$ so that
\begin{eqnarray}
\label{re2}
\fl\frac{\partial \overline{p}(\x,t|\X,0)}{\partial t}
&=- \sum_{j=1}^d\frac{\partial \overline{f}_j(\x,t)\overline{p}(\x,t|\X,0)}{\partial x_j}+D \sum_{j=1}^d\frac{\partial^2\overline{p}(\x,t|\X,0)}{\partial x_j^2}\\
\fl&\quad +\left [\omega(\x|\x_r,T-t)-\delta(\x-\x_r)\int_{\infty}^{\infty} \omega({\bf z}|\x_r,T-t)d{\bf z}\right ]\overline{p}(\x_r,t|\X,0), \nonumber 
\end{eqnarray}
with $\overline{f}_j(\x,t)$ given by equation (\ref{barft}). This describes a Markov diffusion process of the following form: if $\XXo(t)\neq \x_r$ then
\numparts
\begin{eqnarray}
d\Xo_j(t)&=\overline{f}_j(\XXo(t),t)dt+\sqrt{2D}dW_j(t),
\end{eqnarray}
whereas if $\X(t)=\x_r$ then
\begin{eqnarray}
\fl d\Xo_j(t)&=\left \{ \begin{array}{ll}\overline{f}_j(\x_r,t)dt+\sqrt{2D}dW_j(t) & \mbox{ with probability } 1-\frac{\displaystyle rdt}{\displaystyle p(\x_r,T-t)}\\ & \\
x_j-x_{r,j}  &\mbox{ with probability }\frac{\displaystyle p(\x,T-t)rdt}{\displaystyle p(\x_r,T-t)}.\end{array}\right .
\end{eqnarray}
\endnumparts

Finally, taking $p(\x,0)=\rho(\x)$ we have
\numparts
\begin{eqnarray}
d\Xo_j(t)&=\overline{f}_j(\XXo(t))dt+\sqrt{2D}dW_j(t),
\end{eqnarray}
with $\overline{f}_j(\x)$ given by equation (\ref{barf}), whereas if $\X(t)=\x_r$ then
\begin{eqnarray}
\fl d\Xo_j(t)&=\left \{ \begin{array}{ll}\overline{f}_j(\x_r)dt+\sqrt{2D}dW_j(t) & \mbox{ with probability } 1-\frac{\displaystyle rdt}{\displaystyle \rho(\x_r)}\\ & \\
x_j-x_{r,j} &\mbox{ with probability }\frac{\displaystyle \rho(\x)rdt}{\displaystyle \rho(\x_r)}.\end{array}\right .
\end{eqnarray}
\endnumparts
The forward FP equation becomes
\begin{eqnarray}
\frac{\partial \overline{p}(\x,t|\X,0)}{\partial t}
&=- \sum_{j=1}^d\frac{\partial \overline{f}_j(\x)\overline{p}(\x,t|\X,0)}{\partial x_j}+D \sum_{j=1}^d\frac{\partial^2\overline{p}(\x,t|\X,0)}{\partial x_j^2} \nonumber \\
&\quad +\frac{r}{\rho(\x_r)} \left [ \rho(\x)-\delta(\x-\x_r)\right ]\overline{p}(\x_r,t|\X,0).
\label{rFP2}
\end{eqnarray}
Similarly, the backward FP equation for the reverse-time process with stochastic resetting is
\begin{eqnarray}
\fl \frac{\partial \overline{p}(\x,t|\X,0)}{\partial t}
&=  \sum_{j=1}^d\overline{f}_j(\X)\frac{\partial\overline{p}(\x,t|\X,0)}{\partial X_j}+D \sum_{j=1}^d\frac{\partial^2\overline{p}(\x,t|\X,0)}{\partial X_j^2}  \\
\fl &\qquad +\delta(\X-\x_r)\frac{r}{\rho(\x_r)}\left [\int_{\R^d}\rho(\X)\overline{p}(\x,t|\X,0)d\X-\overline{p}(\x,t|\x_r,0)\right ].\nonumber
\end{eqnarray}
Finally, the backward committor functions is obtained from the equation
\begin{equation}
\overline{\mathbb L}_r\overline{q}(\x)=0,\quad \overline{q}(\x)=1 \mbox{ for } \x \in \partial A,\quad \overline{q}(\x)=0\mbox{ for } \x \in \partial B,
\label{rb}
\end{equation}
where
\begin{eqnarray}
\fl \overline{\mathbb L}_r\overline{q}(\x) 
&\equiv  \sum_{j=1}^d\overline{f}_j(\x)\frac{\partial\overline{q}(\x)}{\partial x_j}+D \sum_{j=1}^d\frac{\partial^2\overline{q}(\x)}{\partial x_j^2} +\frac{r\delta(\x-\x_r)}{\rho(\x_r)}\left [\int_{\R^d} \rho(\x)\overline{q}(\x)d\x-\overline{q}( \x_r)\right ].\nonumber \\
\fl
\label{rbFP2}
\end{eqnarray}

\subsection{Calculation of the transition rate $k_{AB}$} In order to calculate the target accumulation rate $k_{AB}$, we need to determine the generalizations of equations (\ref{KAB}) and (\ref{KAB2}) in the presence of resetting. We  proceed by extending the derivations presented in the appendices of Ref. \cite{Metzner06}. A crucial assumption in these derivations is that the underlying stochastic process is ergodic with respect to the stationary density along the lines of equation (\ref{ergodic}). 

\begin{figure}[b!]
\raggedleft
\includegraphics[width=12cm]{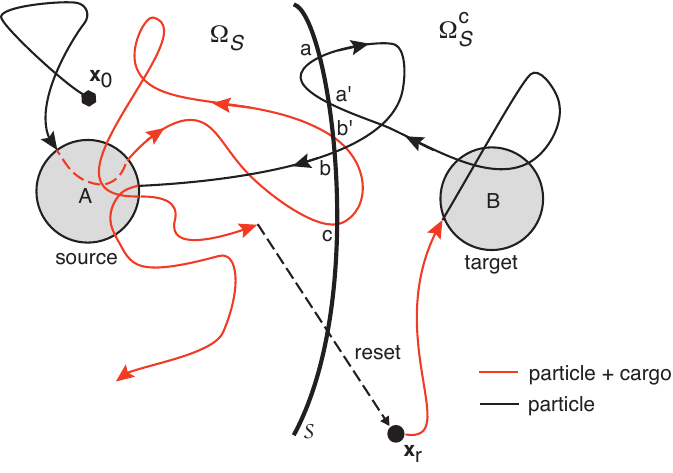}
\caption{Transport process with stochastic resetting. The dividing surface $S \in \Omega_{AB}$ is the boundary of a region $\Omega_S$ with $A\cap \Omega_S=A$ and $B\cap \Omega_S=\emptyset$. The sample trajectory smoothly crosses $S$ from left to right at the points $a,b,c$ and from right to left at the points $a',b'$. The trajectory can also jump across $S$ by resetting from a point $\x\in \Omega_s$ if $\x_r\in \Omega_S^c\equiv \R^d\backslash \Omega_S$ (or from a point $\x\in \Omega_s^c$ if $\x_r\in \Omega_S$ -- not shown).}
\label{fig3}
\end{figure}

It is clear from Fig. \ref{fig3} that there are two distinct ways in which the particle can cross the surface $S$: either continuously, as at the points $a,b,c,a',b'$, or as a jump via stochastic resetting. These two cases will emerge from the analysis. The starting point is an equation equating $k_{AB}$ with the time-avergae of the total probability flux of transport trajectories across $S$ \cite{Metzner06}:
\begin{eqnarray}
k_{AB}&=&\lim_{s\rightarrow 0^+}\frac{1}{s} \lim_{T\rightarrow \infty} \frac{1}{2T}\int_{\calR\cap [-T,T]}\bigg \{\chi_{\Omega_S}(\x(t))\chi_{\Omega_{S}^c}(\x(t+s))\nonumber\\
&&\quad - \chi_{\Omega_{S}^c}(\x(t)\chi_{\Omega_{S}}(\x(t+s))\bigg \}dt.
\end{eqnarray}
Assuming the stochastic process is ergodic, we first take the limit $T\rightarrow \infty$ to obtain
\begin{eqnarray}
k_{AB}&=&\lim_{s\rightarrow 0^+}\frac{1}{s}\int_{\Omega_S}\rho(\y)\overline{q}(\y)\E_{\y}\big [q(\x(s))\chi_{\Omega_{S}^c}(\x(s))\big ]d\y\nonumber\\
&&\quad - \lim_{s\rightarrow 0^+}\frac{1}{s}\int_{\Omega_S^c}\rho(\y)\overline{q}(\y)\E_{\y}\big [q(\x(s))\chi_{\Omega_{S}}(\x(s))\big ]d\y.
\label{exp1}
\end{eqnarray}
The committor functions $\overline{q}(\y)$ and $q(\x(s))$ ensure that we only sum over transport trajectories. In addition $\E_{\y}$ denotes expectation conditional on $\x(0)=\y$. In the case of a smooth function $\phi(\x)$, we have
\begin{eqnarray}
\lim_{t\rightarrow 0^+} \frac{\E_{\y}[\phi(\x(t))]-\phi(\y)}{t}&=
\lim_{t\rightarrow 0^+}\frac{1}{t}\left \{\int_{\R^d}\phi(\x)p(\x,t|\y)d\x-\phi(\y)\right \} \nonumber \\
&=
\lim_{t\rightarrow 0^+}\frac{1}{t}\left \{\int_{\R^d}\phi(\x)[p(\x,t|\y)- p(\x,0|\y)]d\x\right \}\nonumber \\
&=\lim_{t\rightarrow 0}\int_{\R^d}\phi(\x) \frac{\partial p(\x,t|\y)}{\partial t}d\x.
\end{eqnarray}
Substituting for $\partial p/\partial t$ using equation (\ref{fwr}), integrating by parts and using the identity $p(\x,0|\y)=\delta(\y-\x)$ gives
\begin{eqnarray}
\fl \lim_{t\rightarrow 0^+} \frac{\E_{\y}[\phi(\x(t))]-\phi(\y)}{t}&=
\sum_{j=1}^d f_j(\y) \frac{\partial \phi( \y)}{\partial y_j}+D\sum_{j=1}^d\frac{\partial^2\phi(\y)}{\partial y_j^2} -r\phi(\y)+r \phi( \x_r)\nonumber \\ \fl&\equiv {\mathbb L }_{r }\phi(\y).
\label{exp2}
\end{eqnarray}

In order to evaluate the expectations in equation (\ref{exp1}) using equation (\ref{exp2}), we need to regularize the discontinuous indicator functions $\chi_{\Omega_{S}}$ and $\chi_{\Omega_{S}^c}$. Following Ref. \cite{Metzner06}, we introduce a differentiable interfacial function $h_{\delta}$ with 
\begin{equation}
h_{\delta}(\y)=\left \{ \begin{array}{cc} 1 & \mbox{for } \y \in \Omega_s \mbox{ and } \mbox{dist}(\y,S)>\delta \\
0 & \mbox{for } \y \in \Omega_s^c \mbox{ and } \mbox{dist}(\y,S)>\delta \end{array}
\right . ,
\end{equation}
where $\mbox{dist}(\y,S)=\min_{{\bf z}\in S}|\y-{\bf z}|$, and which interpolates smoothly between $0$ and $1$ within the boundary layer of width $2\delta$. It follows that equation (\ref{exp1}) is the limit as $\delta \rightarrow 0$ of
\begin{eqnarray}
I_{\delta}&=\lim_{s\rightarrow 0^+}\frac{1}{s}\int_{\R^d}\rho(\y)\overline{q}(\y)h_{\delta}(\y)\E_{\y}\big [q(\x(s))[1-h_{\delta}(\x(s))]\big ]d\y\nonumber\\
&\quad - \lim_{s\rightarrow 0^+}\frac{1}{s}\int_{\R^d}\rho(\y)\overline{q}(\y)[1-h_{\delta}(\y)]\E_{\y}\big [q(\x(s))h_{\delta}(\x(s))\big ]d\y.
\end{eqnarray}
Using the result
\begin{equation} 
\lim_{\x\rightarrow \y} \big\{ h_{\delta}(\y)q(\x)[1-h_{\delta}(\x)]-
[1-h_{\delta}(\y)]q(\x) h_{\delta}(\x)\big \}=0,
\end{equation}
we can now apply equation (\ref{exp2}) to yield
\begin{eqnarray}
 I_{\delta}&= \int_{\R^d}\rho(\y)\overline{q}(\y)h_{\delta}(\y){\mathbb L}_{r}\big [q(\y)[1-h_{\delta}(\y)]\big ]d\y \nonumber \\
 &\quad -  \int_{\R^d}\rho(\y)\overline{q}(\y)[1-h_{\delta}(\y)]{\mathbb L}_{r}\big [q(\y)h_{\delta}(\y)\big ]d\y\nonumber \\
 &=-\int_{\R^d}\rho(\y)\overline{q}(\y){\mathbb L}_{r}\big [q(\y)h_{\delta}(\y)\big ]d\y ,
\end{eqnarray}
since ${\mathbb L}_rq(\y)=0$. If we now substitute the explicit form for ${\mathbb L}_{r}$, see equation (\ref{bwr}), we find that
\begin{eqnarray}
\fl I_{\delta}&=- \int_{\R^d}\rho(\y)\overline{q}(\y)\bigg \{q(\y)\sum_{j=1}^d f_j(\y) \frac{\partial  h_{\delta}(\y)}{\partial y_j}+D\sum_{j=1}^d\bigg [q(\y)\frac{\partial^2h_{\delta}(\y)}{\partial y_j^2}+2\frac{\partial  h_{\delta}(\y)}{\partial y_j}\frac{\partial  q(\y)}{\partial y_j}\bigg ]\nonumber \\
\fl &\quad -  rq(\x_r)h_{\delta}(\y) +rq(\x_r)h_{\delta}(\x_r) \bigg\}d\y .
\end{eqnarray}
Finally, integrating by parts the term involving the second order derivative of $h_{\delta}$ gives
\begin{eqnarray}
\fl I_{\delta}&=- \int_{\R^d}\sum_{j=1}^d \frac{\partial  h_{\delta}(\y)}{\partial y_j}\bigg \{\overline{q}(\y)q(\y)J_j(\y)+D\rho(\y)\overline{q}(\y) \frac{\partial  q(\y)}{\partial y_j}-D\rho(\y)q(\y) \frac{\partial  \overline{q}(\y)}{\partial y_j}\bigg \}d\y\nonumber \\
\fl &\quad +rq(\x_r)\int_{\R^d}\rho(\y)\overline{q}(\y)\bigg\{ h_{\delta}(\y) -h_{\delta}(\x_r) \bigg\}d\y ,\label{exp3}
\end{eqnarray}
where $J_j$ is the current defined in equation (\ref{fwrstat}). 

Recall that for any sutiably defined vector field ${\bf v}(\y)=(v_1(\y),\ldots v_d(\y))^{\top}$, we have
\begin{eqnarray*}
\fl \lim_{\delta \rightarrow 0} \int_{\R^d}\sum_{j=1}^d \frac{\partial  h_{\delta}(\y)}{\partial y_j}v_j(\y)d\y&=-\lim_{\delta \rightarrow 0} \int_{\R^d}h_{\delta}(\y)\sum_{j=1}^d \frac{\partial v_j(\y) }{\partial y_j}d\y\nonumber \\
\fl &=-\int_{\Omega_S}\sum_{j=1}^d \frac{\partial v_j(\y) }{\partial y_j}d\y=-\int_S{\bf v}(\x) \cdot \n(\x)d\sigma(\x).
\end{eqnarray*}
We have used integration by parts, the definition of $h_{\delta}(\y)$, and the divergence theorem. Hence, taking the limit $\delta \rightarrow 0$ in equation (\ref{exp3}) yields 
\begin{eqnarray}
\label{KABr}
\fl k_{AB}&=\lim_{\delta \rightarrow 0}I_{\delta} \\
\fl &=\int_S{\bf J}_{AB}(\x) \cdot \n(\x)d\sigma(\x)+r  q(\x_r)\left [\int_{\Omega_S}\rho(\y)\overline{q}(\y)   d\y- \chi_{\Omega_S}(\x_r)\int_{\R^d}\rho(\y)\overline{q}(\y)   d\y\right ] ,
\nonumber
\end{eqnarray}
where 
\begin{eqnarray}
\fl J_{AB,j} (\x) =q (\x)\overline{q} (\x) J_j (\x)+D\overline{q} (\x)\rho (\x) \frac{\partial q (\x)}{\partial x_j}-D q (\x)\rho (\x) \frac{\partial \overline{q} (\x)}{\partial x_j}.
\label{JABfull}
\end{eqnarray}
Equation (\ref{JABfull}) has the same form as equation (\ref{JAB}) for the current ${\bf J}_{AB,j}^{(0)}$ without resetting, in which the triplet $(\rho_0,q_0,\overline{q}_0)$ is replaced by $(\rho,q,\overline{q})$. Moreover, the NESS $\rho(\x)$ is  given by equation (\ref{sas}), while the committor functions $q(\x)$ and $\overline{q}(\x)$ satisfy equations (\ref{rf}) and (\ref{rb}), respectively. It follows that ${\bf J}_{AB}(\x)$ is an implicit function of the resetting rate $r$. The first term on the right-hand side of equation (\ref{KABr}) represents the contribution from smooth paths crossing $S$, see Fig. \ref{fig3}. The two additional terms on the right-hand side of equation (\ref{KABr}) take into account paths that jump across $S$ due to resetting. In particular, the contribution to $k_{AB}$ due to resetting can be decomposed according to
\[  \fl r  q(\x_r) \int_{\Omega_S}\rho(\y)\overline{q}(\y)   d\y\mbox{ for } \x_r \in \Omega_S^c \quad \mbox{ and } \quad -r  q(\x_r) \int_{\Omega_S^c}\rho(\y)\overline{q}(\y)   d\y \mbox{ for } \x_r \in \Omega_S. \]

We now need to check that $k_{AB}$ is independent of the dividing surface $S$. In the absence of stochastic resetting, the current ${\bf J}_{AB}^{(0)}$ is divergence-free and the result follows immediately from equation (\ref{KAB}) and (\ref{JAB}). Let us determine the divergence of the corresponding current ${\bf J}_{AB}$ in the presence of resetting:
\begin{eqnarray*}
\fl \sum_{j=1}^d\frac{\partial J_{AB,j}(\x)}{\partial x_j}&=
\overline{q}(\x) \sum_{j=1}^d\bigg \{ \bigg [f_j(\x)\rho(\x) -D \frac{\partial \rho(\x)}{\partial x_j}\bigg ]\frac{\partial q(\x)}{\partial x_j}+D \rho(\x) \frac{\partial^2 q(\x)}{\partial x_j^2}\\
\fl &\hspace{1cm}+D\frac{\partial \rho(\x)}{\partial x_j}\frac{\partial q(\x)}{\partial x_j}\bigg \}+q(\x)\overline{q}(\x)\sum_{j=1}^d  \frac{\partial}{\partial x_j}\bigg [f_j(\x)\rho(\x) -D \frac{\partial \rho(\x)}{\partial x_j}\bigg ]\\
\fl &\quad +q(\x)\sum_{j=1}^d \bigg \{\bigg [f_j(\x)\rho(\x) -2D \frac{\partial \rho(\x)}{\partial x_j}\bigg ]\frac{\partial \overline{q}(\x)}{\partial x_j}-D\rho(\x)\frac{\partial^2 \overline{q}(\x)}{\partial x_j^2}\bigg\}.
\end{eqnarray*}
Using equations (\ref{fwr}), (\ref{rf}) and (\ref{rb}), this reduces to
\begin{eqnarray}
\fl\sum_{j=1}^d\frac{\partial J_{AB,j}(\x)}{\partial x_j}&=
r\rho(\x)\overline{q}(\x) [q(\x)-q(\x_r)]+rq(\x)\overline{q}(\x)[\delta(\x-\x_r)-\rho(\x)]\nonumber \\
\fl&\quad +\rho(\x)q(\x)\delta(\x-\x_r)\frac{r}{\rho(\x_r)}\left [\int_{\R^d}\rho(\x)\overline{q}(\x)d\x-\overline{q}( \x_r)\right ]\nonumber\\
\fl &=-rq(\x_r)\bigg \{\rho(\x)\overline{q}(\x)+\delta(\x-\x_r) \int_{\R^d}\rho(\x)\overline{q}(\x)d\x \bigg\}.\label{divJ}
\end{eqnarray}

\begin{figure}[b!]
\centering
\includegraphics[width=8cm]{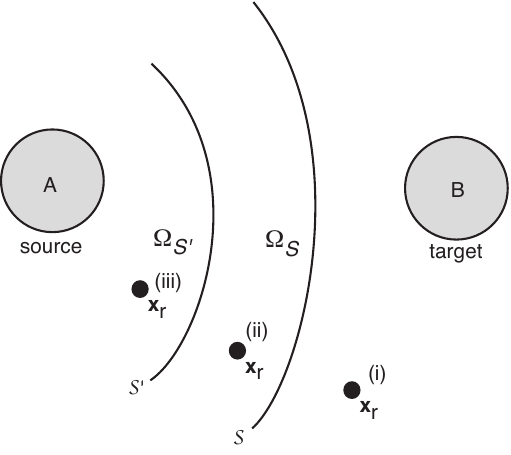}
\caption{Two concentric domains $\Omega_S$ and $\Omega_{S'}$, $\Omega_{S'}\subset \Omega_S$, with corresponding dividing surfaces $S$ and $S'$. Also shown are three different positions of the reset point $\x_r$ relative to these surfaces.}
\label{fig4}
\end{figure}

Consider two concentric domains $\Omega_S$ and $\Omega_{S'}$, $\Omega_S'\subset \Omega_S$ with corresponding dividing surfaces $S$ and $S'$. Denote the corresponding accumulation rates by $k_{AB}(S)$ and $k_{AB}(S')$. In order to compare the two rates, it is necessary to specify the location of $\x_r$ with respect to the two domains, see Fig. \ref{fig5}. First, suppose that $\x_r\in \Omega_S^c$ (case (i)). Equation (\ref{KABr}) implies that
\begin{eqnarray}
\fl k_{AB}(S)-k_{AB}(S')&=\int_{S-S'}{\bf J}_{AB}(\x) \cdot \n(\x)d\sigma(\x)+r  q(\x_r) \int_{\Omega_S-\Omega_{S'}}\rho(\y)\overline{q}(\y)   d\y\nonumber \\
\fl&=\int_{\Omega_S-\Omega_S'}\sum_{j=1}^d\frac{\partial J_{AB,j}(\x)}{\partial x_j}d\x+r  q(\x_r) \int_{\Omega_S-\Omega_{S'}}\rho(\y)\overline{q}(\y)   d\y\nonumber \\
\fl&=0.
\end{eqnarray}
The last line follows from equation (\ref{divJ}) and the assumption that $\x_r \notin \Omega_S-\Omega_{S'}$, which implies that only the first term on the right-hand side of (\ref{divJ}) contributes. On the other hand, if $\x_r\in \Omega_S-\Omega_{S'}$ (case (ii)), then
\begin{eqnarray}
\fl k_{AB}(S)-k_{AB}(S')&=\int_{\Omega_S-\Omega_S'}\sum_{j=1}^d\frac{\partial J_{AB,j}(\x)}{\partial x_j}d\x+r  q(\x_r) \int_{\Omega_S-\Omega_{S'}}\rho(\y)\overline{q}(\y)   d\y\nonumber \\
\fl &\quad -r  q(\x_r) \int_{\R^d}\rho(\y)\overline{q}(\y)   d\y=0.
\end{eqnarray}
The additional integral over $\R^d$ is cancelled by the term involving the Dirac delta function in equation (\ref{divJ}). Finally, it can be checked that 
$k_{AB}(S)=k_{AB}(S')$ when $\x\in \Omega_{S'}$ (case (iii)). We conclude that although the probability current ${\bf J}_{AB}$ is not divergence-free in the presence of stochastic resetting, the resulting rate $k_{AB}$ is independent of the dividing surface $S$.

Having obtained the resetting-dependent version of equation (\ref{KAB}), we now turn to the analog of equation (\ref{KAB2}). Following Ref. \cite{Metzner06}, we introduce the forward iso-committor surface $\calM(\xi)=\{\x\in \R^d: q(\x)=\xi\}$ with $\xi\in [0,1]$ and define the integral
\begin{equation}
\calA(\xi)=D\int_{\calM(\xi)}\rho(\x)\sum_{j=1}^d n_j(\x)\frac{\partial q(\x)}{\partial x_j}d\sigma(\x)  .
\end{equation}
Note that $\calM(0)=\partial A$, where $\partial A$ is the surface of the source domain. Since $q(\x)=0$ and $\overline{q}(\x)=1$ on $\partial A$, it follows from equation (\ref{JABfull}) that
\begin{eqnarray}
J_{AB,j}(\x)=D\rho(\x) \frac{\partial q(\x)}{\partial x_j},\quad \x\in \partial A,
\end{eqnarray}
and hence
\begin{equation}
\calA(0)=\int_{\partial A}{\bf J}_{AB}(\x) \cdot \n(\x)d\sigma(\x)=k_{AB}-r  q(\x_r) \int_{A}\rho(\y) d\y.
\label{calAi}
\end{equation}
We have also used equation (\ref{KABr}) with $\Omega_S=A$.
In addition, rewriting $A(\xi)$ as
\begin{equation}
A(\xi)=D\int_{\R^d}\rho(\x)\sum_{j=1}^d  \left (\frac{\partial q(\x)}{\partial x_j}\right )^2\delta(q(\x)-\xi)d\x,
\label{Axio}
\end{equation}
we find
\begin{eqnarray*}
\frac{d\calA(\xi)}{d\xi}&=-D\int_{\R^d}\rho(\x)\sum_{j=1}^d  \left (\frac{\partial q(\x)}{\partial x_j}\right )^2\delta'(q(\x)-\xi)d\x,\nonumber \\
&=-D\int_{\R^d}\rho(\x)\sum_{j=1}^d   \frac{\partial q(\x)}{\partial x_j}\frac{\partial}{\partial x_j} \delta(q(\x)-\xi)d\x\nonumber \\
 &=D\int_{\R^d} \sum_{j=1}^d \bigg \{\rho(\x) \frac{\partial^2 q(\x)}{\partial x_j^2} +\frac{\partial  q(\x)}{\partial x_j } \frac{\partial  \rho(\x)}{\partial x_j }\bigg \}\delta(q(\x)-\xi)d\x      \\
 &=\int_{\R^d}  \bigg \{\sum_{j=1}^d\bigg[-\rho(\x) f_j(\x)\frac{\partial q(\x)}{\partial x_j} +D\frac{\partial  q(\x)}{\partial x_j } \frac{\partial  \rho(\x)}{\partial x_j }\bigg ]\\
 &\quad +r\rho(\x)[q(\x)-q(\x_r)] \bigg \}\delta(q(\x)-\xi)d\x   .
\end{eqnarray*}
We have performed an integration by parts and used equation (\ref{rf}).
Using equation (\ref{fwrstat}), we can combine the two terms in square brackets to give
\begin{eqnarray*}
\fl-\int_{\R^d} \sum_{j=1}^d   J_j(\x)\frac{\partial q(\x)}{\partial x_j} \delta(q(\x)-\xi)d\x  &= -\int_{\calM(\xi)} \sum_{j=1}^d  n_j(\x) J_j(\x)d\sigma(\x)\\
\fl &=-\int_{\Omega(\xi)} \nabla \cdot{\bf J}(\x)d\x \\
&=r\int_{\Omega(\xi)} [\rho(\x)-\delta(\x-\x_r)]d\x ,
\end {eqnarray*}
where $\Omega(\xi)$ is the domain containing $A$ whose surface is $\calM(\xi)$. Hence,
\begin{eqnarray}
\fl \frac{d\calA(\xi)}{d\xi}
 &=\int_{\R^d}   r\rho(\x)[q(\x)-q(\x_r)] 
 \delta(q(\x)-\xi)d\x +r\int_{\Omega(\xi)} [\rho(\x)-\delta(\x-\x_r)]d\x. 
 \label{calAd}
\end{eqnarray}

In order to integrate equation (\ref{calAd}) with respect to $\xi$, we make use of the following identities. First, for any integrable function $\phi(\x)$, we have
\begin{eqnarray}
\int_{\Omega(\xi) }\phi(\x)d\x=\int_0^{\xi} d\xi''\int_{\R^d}\phi(\x)\delta(q(\x)-\xi'')d\x +\int_{  A}\phi(\x)d\x.
\label{dint0}
\end{eqnarray}
Second, integrating both sides with respect to $\xi$ implies that
\begin{eqnarray}
\fl \int_0^{\xi}d\xi' \int_{\Omega(\xi') }\phi(\x)d\x&=\int_0^{\xi}d\xi' \int_0^{\xi'} d\xi''\int_{\R^d}\phi(\x)\delta(q(\x)-\xi'')d\x +\xi \int_{  A}\phi(\x)d\x\nonumber \\
&=\int_0^{\xi}d\xi'' \int_{\xi''}^{\xi} d\xi'\int_{\R^d}\phi(\x)\delta(q(\x)-\xi'')d\x +\xi \int_{  A}\phi(\x)d\x\nonumber \\
&=\int_0^{\xi}d\xi''  \int_{\R^d}[\xi-q(\x)]\phi(\x)\delta(q(\x)-\xi'')d\x +\xi \int_{  A}\phi(\x)d\x\nonumber \\
&=\int_{\Omega(\xi)\backslash A }[\xi-q(\x)]\phi(\x)d\x+\xi \int_{  A}\phi(\x)d\x.
\label{dint}
\end{eqnarray}
The last line follows from another application of equation (\ref{dint0}). 
Now integrating both sides of equation (\ref{calAd}) with respect to $\xi$ using equations (\ref{dint0}) and (\ref{dint}) gives
\begin{eqnarray}
\fl\calA(\xi)&=\calA(0)+r\int_{\Omega(\xi)\backslash A}\rho(\x)[q(\x)-q(\x_r)]d\x \nonumber \\
\fl&\quad +r\int_{\Omega(\xi)\backslash A} [\xi-q(\x)][\rho(\x)-\delta(\x-\x_r)]d\x+r\xi \int_{  A}[\rho(\x)-\delta(\x-\x_r)]d\x\nonumber \\
\fl &=k_{AB} +r\int_{\Omega(\xi)}\rho(\x)[q(\x)-q(\x_r)] d\x+r\int_{\Omega(\xi)} [\xi-q(\x)][\rho(\x)-\delta(\x-\x_r)]d\x.\nonumber  \\
\fl &=k_{AB}+r[\xi-q(\x_r)]\int_{\Omega(\xi)}[\rho(\x)-\delta(\x-\x_r)]d\x.
\label{Axi}
\end{eqnarray}
We have also used equation (\ref{calAi}) and $q(\x)=0$ for all $\x\in \overline{A}$.
Finally, integrating both sides of equation (\ref{Axi}) with respect to $\xi$ and using equation (\ref{Axio}) leads to the result
\begin{eqnarray}
  k_{AB}&=D\int_{\R^d\backslash A\cup B}\rho(\x)\sum_{j=1}^d  \left (\frac{\partial q(\x)}{\partial x_j}\right )^2d\x\nonumber \\
  &\quad -r\int_0^1[\xi-q(\x_r)]\left [\int_{\Omega(\xi)}[\rho(\x) -\delta(\x-\x_r)] d\x \right ]d\xi.  \end{eqnarray}
We evaluate the double integral on the second line using the identity (\ref{dint0}):
\begin{eqnarray}
\fl \mbox{second line}=-r\left [I_{\xi}+\left (\frac{1}{2}-q(\x_r)\right )\int_{  A}\rho(\x)  d\x\right ],
\end{eqnarray}
where
\begin{eqnarray}
\fl I_{\xi}&\equiv \int_0^1d\xi [\xi-q(\x_r)]\left [\int_0^{\xi} d\xi''\int_{\R^d}[\rho(\x) -\delta(\x-\x_r)] \delta(q(\x)-\xi'')d\x  \right ] \nonumber\\
\fl &=\int_0^1d\xi ''\int_{\xi''}^1d\xi [\xi-q(\x_r)]\left [ \int_{\R^d}[\rho(\x) -\delta(\x-\x_r)] \delta(q(\x)-\xi'')d\x\right ]\nonumber \\
\fl &\quad \nonumber\\
\fl &=\int_0^1d\xi ''\left [\frac{1-{\xi''}^2}{2}-(1-\xi'')q(\x_r)\right ]\left [ \int_{\R^d}[\rho(\x) -\delta(\x-\x_r)] \delta(q(\x)-\xi'')d\x\right ]\nonumber \\
\fl &=\int_0^1d\xi ''\left [ \int_{\R^d}\left [\frac{1-q(\x)^2}{2}-(1-q(\x))q(\x_r)\right ][\rho(\x) -\delta(\x-\x_r)] \delta(q(\x)-\xi'')d\x\right ]\nonumber \\
\fl &= \left [ \int_{\R^d\backslash A\cup B}\left [\frac{1-q(\x)^2}{2}-(1-q(\x))q(\x_r)\right ][\rho(\x) -\delta(\x-\x_r)] d\x\right ]\nonumber \\
\fl &=-\frac{(1-q(\x_r))^2}{2}-  \int_{\R^d\backslash A\cup B}(1-q(\x))\left [-\frac{1+q(\x)}{2}+q(\x_r)\right ]\rho(\x) d\x.
\end{eqnarray}
Combining our various results, we have
\begin{eqnarray}
 \fl  k_{AB}&= \int_{\R^d\backslash A\cup B}\rho(\x)\bigg \{D\sum_{j=1}^d  \left (\frac{\partial q(\x)}{\partial x_j}\right )^2+r(1-q(\x))\left [-\frac{1+q(\x)}{2}+q(\x_r)\right ]\bigg \}d\x\nonumber \\
  \fl &\quad +r\frac{(1-q(\x_r))^2}{2}-r\left [\frac{1}{2}-q(\x_r) \right ]\int_{  A}\rho(\x)  d\x.\end{eqnarray}
Rearranging various terms, we have
\begin{eqnarray}
 \fl  k_{AB}&= \int_{\R^d\backslash A\cup B}\rho(\x)\bigg \{D\sum_{j=1}^d  \left (\frac{\partial q(\x)}{\partial x_j}\right )^2+r\left [\frac{q(\x)^2-q(\x_r)^2}{2}+(1-q(\x))q(\x_r)\right ]\bigg \}d\x\nonumber \\
  \fl &\quad +r\frac{(1-q(\x_r))^2}{2}\left [1-\int_{\R^d\backslash A\cup B}\rho(\x)-\int_{  A}\rho(\x)  d\x\right ]\nonumber \\
  \fl &\quad +r\frac{q(\x_r)^2}{2} \int_{  A}\rho(\x)  d\x.
  \label{final}\end{eqnarray}
It can be checked that all terms on the right-hand side  of equation (\ref{final}) are positive. Equation (\ref{final}) is the resetting analog of
equation (\ref{KAB2}).

\section{Diffusion with resetting in the interval}

\begin{figure}[b!]
\centering
\includegraphics[width=12cm]{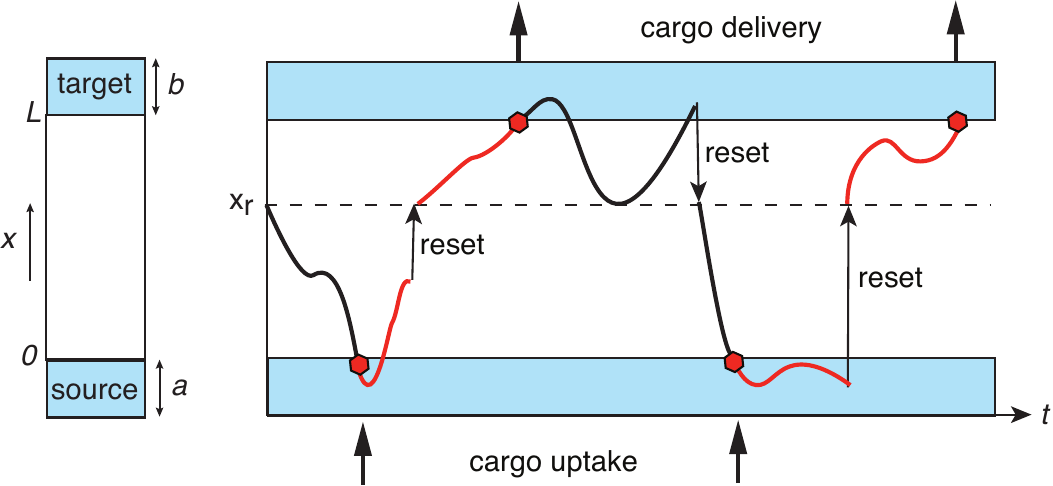}
\caption{Diffusion with resetting in the interval $\Omega=[-a,L+b]$ with source domain $A=[-a,0]$ and target domain $B=[L,L+b]$.}
\label{fig5}
\end{figure}

As an illustrative example of the analysis developed in section 3, consider pure diffusion in the finite interval $\Omega= [-a,L+b]$, $a,b>0$, with reflecting boundaries at $x=-a$ and $x=L+b$. Take the source and target domains to be $A=[-a,0]$ and $B=[L,L+b]$, respectively, see Fig. \ref{fig5}(a). We assume that the particle resets to a point $x_r \in (0,L)$ at a rate $r$, see Fig. \ref{fig5}(b). As in Fig. \ref{fig1}(b), the particle carries cargo along sections of the trajectory that link $A$ to $B$.

In order to determine the NESS $\rho(x)$, we consider the Laplace transformed diffusion equation in $[-a,L+b]$ without resetting:
\begin{equation}
\fl D\frac{d^2\widetilde{p}_0(x,s|x_0,0)}{dx^2}-s\widetilde{p}_0(x,s|x_0,0)=-\delta(x-x_0),\quad x\in [-a,L+b].
\end{equation}
This is supplemented by the reflecting boundary conditions
\begin{equation}
\left .\frac{d\widetilde{p}_0(x,s|x_0,0)}{dx}\right|_{x=-a}=0=\left .\frac{d\widetilde{p}_0(x,s|x_0,0)}{dx}\right|_{x=L+b}.
\end{equation}
We can identify $\widetilde{p}_0(x,s|x_0)$ as a Green's function of the modified Helmholtz equation on $[-a,L+b]$. Imposing continuity of $\widetilde{p}(x,s|x_0)$ across $x_0$ and matching the discontinuity in the first derivative yields the solution
 \begin{eqnarray}
 \widetilde{p}_0(x,s|x_0)= \left \{ \begin{array}{cc} \Gamma \widetilde{p}_<(x,s)\widetilde{p}_>(x_0,s), & -a\leq x\leq x_0\\ & \\
 \Gamma\widetilde{p}_>(x,s)\widetilde{p}_<(x_0,s), & x_0\leq x\leq L+b\end{array}
 \right . ,
 \label{solVN}
 \end{eqnarray}
 with
 \begin{equation}
  \Gamma=\Gamma(s)\equiv \frac{1} {\sqrt{sD}\sinh(\sqrt{sD}[L+b+a])},
  \end{equation}
  and
 \begin{equation}
 \fl \widetilde{p}_<(x,s)=\cosh(\sqrt{s/D}[x+a]),\quad \widetilde{p}_>(x,s)= \cosh(\sqrt{s/D}(L+b-x)).
 \end{equation}
  Finally, it follows from equation (\ref{sas}) that $\rho(x)=r\widetilde{p}_0(x,r|x_r)$.

The backward Kolmogorov equation for $q(x)$ is
\begin{eqnarray}
\label{U0q}
  D\frac{d^2q(x)}{d x^2} -rq(x)+r q(x_r)=0,\quad q(0)=0,\quad q(L)=1.
\end{eqnarray}
We also assume that $0<x_r<L$. 
Let $u(x)=q(x)-q(x_r)$ with
\begin{eqnarray}
\label{U0q2}
  D\frac{d^2u(x)}{dx^2} -ru(x)=0,\quad u(0)=-q(x_r),\quad u(L)=1-q(x_r).
\end{eqnarray}
The solution for $u(x)$, $x\in [0,L]$, that satisfies the boundary condition at $x=0$ takes the form
\begin{equation}
u(x)=U\sinh(\alpha x)-q(x_r)\cosh(\alpha x),\quad \alpha=\sqrt{r/D},
\end{equation}
and thus
\begin{equation}
q(x)=U\sinh(\alpha x)+q(x_r)[1-\cosh(\alpha x)],\quad x\in [0,L].
\end{equation}
Setting $x=x_r$ gives the self-consistency condition
\[q(x_r)=U\sinh(\alpha x_r)+q(x_r)[1-\cosh(\alpha x_r)],\]
which yields the relation
\begin{equation}
U=\frac{q(x_r)\cosh(\alpha x_r)}{\sinh(\alpha x_r)}.
\end{equation}
Hence,
\begin{eqnarray}
q(x)&=q(x_r)\left [\frac{ \cosh(\alpha x_r)}{\sinh(\alpha x_r)}\sinh(\alpha x)-\cosh(\alpha x)\right ]+q(x_r)\nonumber \\
&=q(x_r)\left [\frac{ \sinh(\alpha(x-x_r))}{\sinh(\alpha x_r )}+1\right ].
\end{eqnarray}
Finally, imposing the right-hand boundary condition implies that
\[1=q(x_r)\left [\frac{ \sinh(\alpha(L-x_r))}{\sinh(\alpha x_r)}+1\right ],\]
which can be rearranged to give
\begin{equation}
q(x_r)=\frac{\sinh(\alpha x_r)}{\sinh(\alpha x_r )+\sinh(\alpha(L-x_r))}.
\end{equation}
This recovers the result previously derived in Ref. \cite{Pal19} using a different method. Moreover,
\begin{equation}
q(x)=\frac{\sinh(\alpha x_r )+\sinh(\alpha (x-x_r))}{\sinh(\alpha x_r)+\sinh(\alpha(L-x_r))}.
\end{equation}

\begin{figure}[t!]
\centering
\includegraphics[width=10cm]{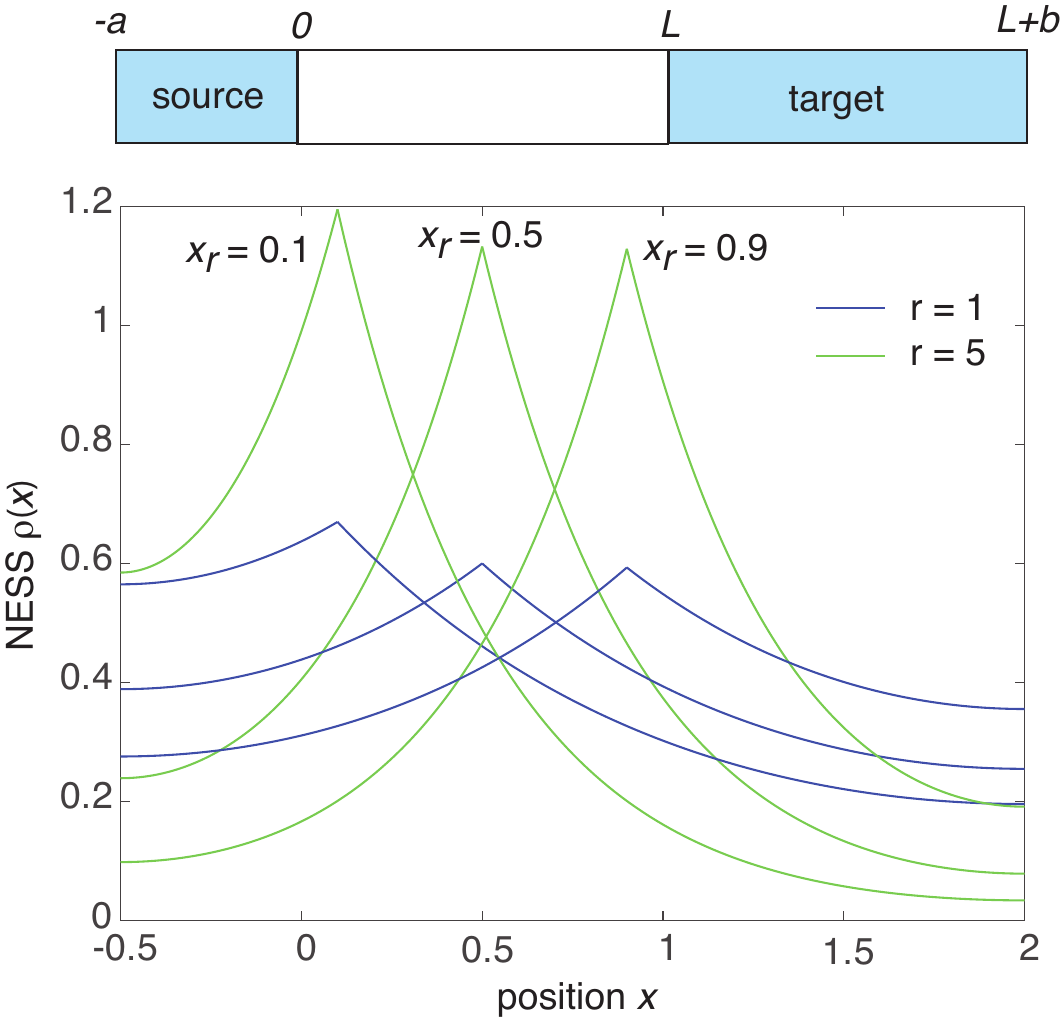}
\caption{Example plots of the NESS in the domain $[-a,L+b]$ with $a=0.5,b=1.0,L=1.0$ and various values of the reset point $x_r$ and rate $r$. We also set $D=1$}
\label{fig6}
\end{figure}

\begin{figure}[t!]
\centering
\includegraphics[width=10cm]{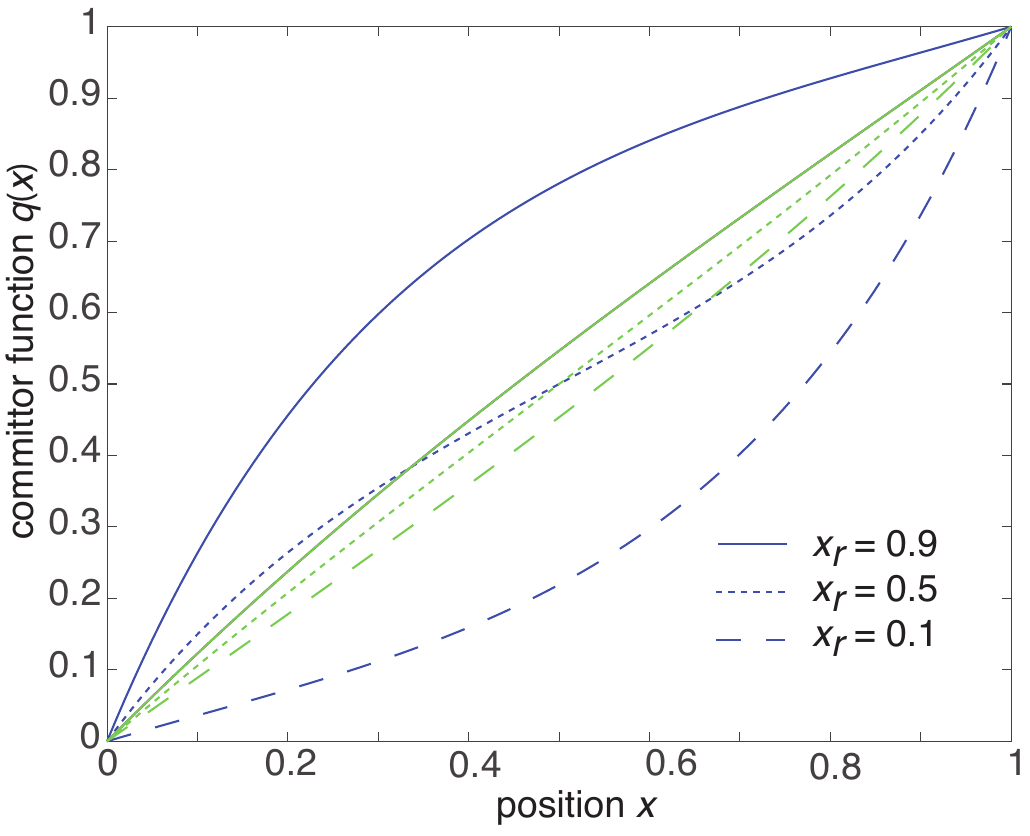}
\caption{Example plots of the forward committor function $q(x)$, $x\in [0,L]$, for $L=1$, $D=1$ and various values of the reset point $x_r$. Blue (green) curves correspond to $r=10$ ($r=1$).}
\label{fig7}
\end{figure}

In Fig. \ref{fig6} we show example plots of the NESS $\rho(x)$, $x\in [-a,L+b]$ for various values of the reset position $x_r$ and the reset rate $r$. As expected there is a cusp at $x=x_r$ whose value increases with $r$. In Fig. \ref{fig7} we show analogous plots of the forward committor function $q(x)$. Note that in the zero resetting limit $r\rightarrow 0$, we have
\begin{equation}
\rho(x)\rightarrow \frac{1}{L+b+a},\ x\in [-a,L+b] \mbox{  and  } q(x) \rightarrow \frac{x}{L},\ x\in [0,L].
\end{equation}
Since the accumulation rate $k_{AB}$ can be expressed solely in terms of $\rho(x)$ and $q(x)$, see equation (\ref{final}), we do not explicitly calculate the backward committor function $\overline{q}(x)$.

The 1D version of equation (\ref{final}) for $k_{AB}$  is
\begin{eqnarray}
 \fl  k_{AB}&= \int_{0}^L\rho(x)\bigg \{D   \left (\frac{\partial q(x)}{\partial x}\right )^2+r\left [\frac{q(x)^2-q(x_r)^2}{2}+(1-q(x))q(x_r)\right ]\bigg \}dx\nonumber \\
  \fl &\quad +r\frac{(1-q(x_r))^2}{2}\left [1-\int_{-a}^L\rho(x) dx\right ]+r\frac{q(x_r)^2}{2} \int_{-a}^0\rho(x)  dx.
  \label{final1D}\end{eqnarray}
The term in curly brackets can be written as
\begin{eqnarray}
 I_1(x) &= \frac{1}{[\sinh(\alpha x_r)+\sinh(\alpha(L-x_r))]^2}
\bigg \{\alpha^2 D\cosh^2\alpha(x-x_r)\nonumber\\
  &\quad +\frac{r}{2}\bigg (2\sinh(\alpha x_r )+\sinh(\alpha (x-x_r))\bigg )\sinh(\alpha (x-x_r)\nonumber \\
  &\quad +r\bigg (\sinh(\alpha [L-x_r])-\sinh(\alpha(x-x_r))\bigg )\sinh(\alpha x_r)\bigg \}.
\end{eqnarray}
Using the fact that $0<x_r<L$, we have
\begin{eqnarray}
\fl k_{AB}&=r\Gamma(r)\widetilde{p}_>(x_r,r)\int_0^{x_r}\widetilde{p}_<(x,r)I_1(x)dx+r\Gamma(r)\widetilde{p}_<(x_r,r)\int_{x_r}^L\widetilde{p}_>(x,r)I_1(x)dx\nonumber \\
\fl &\qquad +\frac{r^2\widetilde{p}_<(x_r,r)\Gamma(r)\sinh^2(\alpha(L-x_r))}{2[\sinh(\alpha x_r)+\sinh(\alpha(L-x_r))]^2}\int_{L}^{L+b}\widetilde{p}_>(x,r)dx\nonumber \\
\fl &\qquad + \frac{r^2\widetilde{p}_>(x_r,r)\Gamma(r)\sinh^2(\alpha x_r)}{2[\sinh(\alpha x_r)+\sinh(\alpha(L-x_r))]^2}\int_{-a}^{0}\widetilde{p}_<(x,r)dx.
\end{eqnarray}

\begin{figure}[t!]
\centering
\includegraphics[width=10cm]{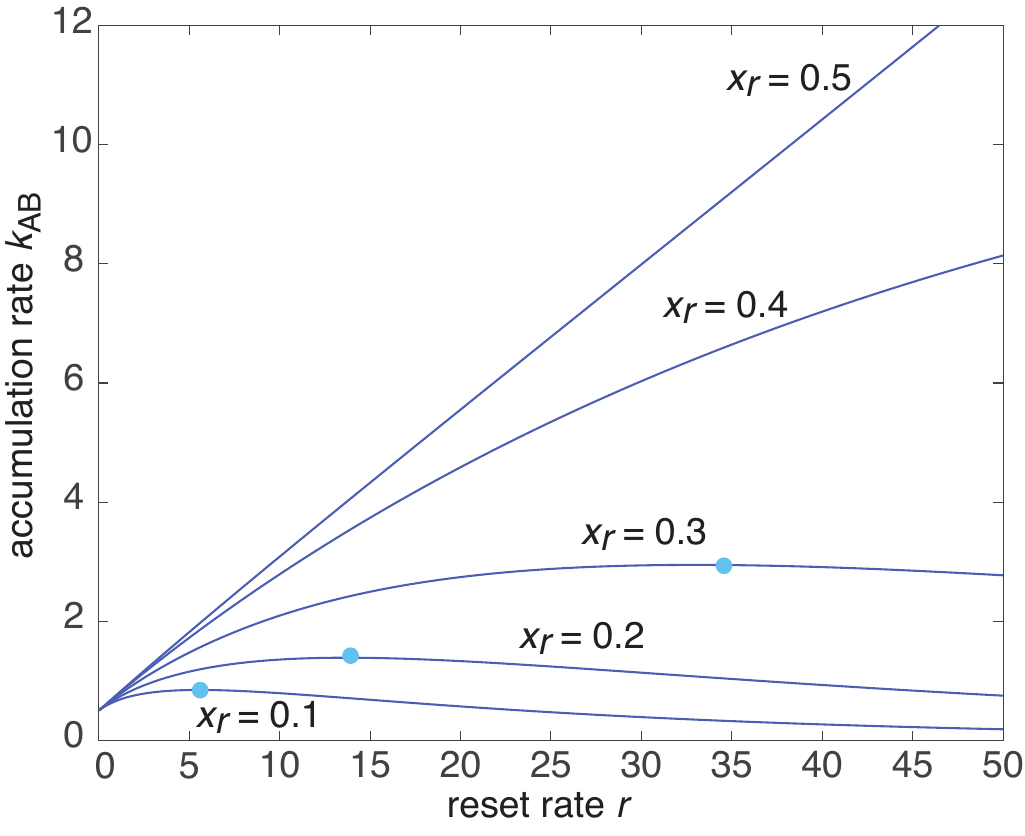}
\caption{Plot of target accumulation rate $k_{AB}$ as a function of the reset rate $r$ for various values of the reset point $x_r$. Other parameters are $L=1$, $a=b=0.5$ and $D=1$. There exists an optimal resetting rate for all $x_r\neq 0.5$. (The curves are invariant under the mapping $x_r\rightarrow 1-x_r$. The maximum when $x_r=0.4,0.6$ occurs at $r\approx 140$ - not shown.)}
\label{fig8}
\end{figure}

\begin{figure}[t!]
\centering
\includegraphics[width=10cm]{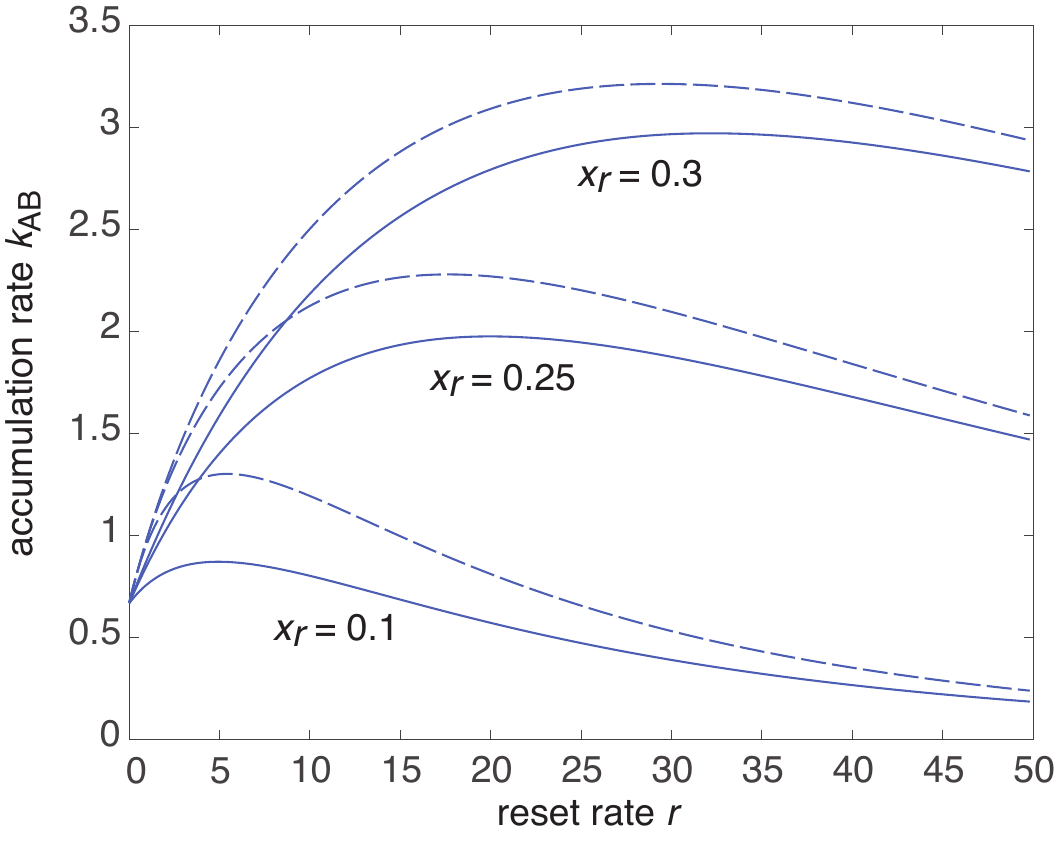}
\caption{Plot of target accumulation rate $k_{AB}$ as a function of the reset rate $r$ for various values of the reset point $x_r$. Other parameters are $L=1$, $a=0.5$, $b=0$, and $D=1$. Dashed curves correspond to the case $1-x_r$.}
\label{fig9}
\end{figure}

\begin{figure}[t!]
\centering
\includegraphics[width=10cm]{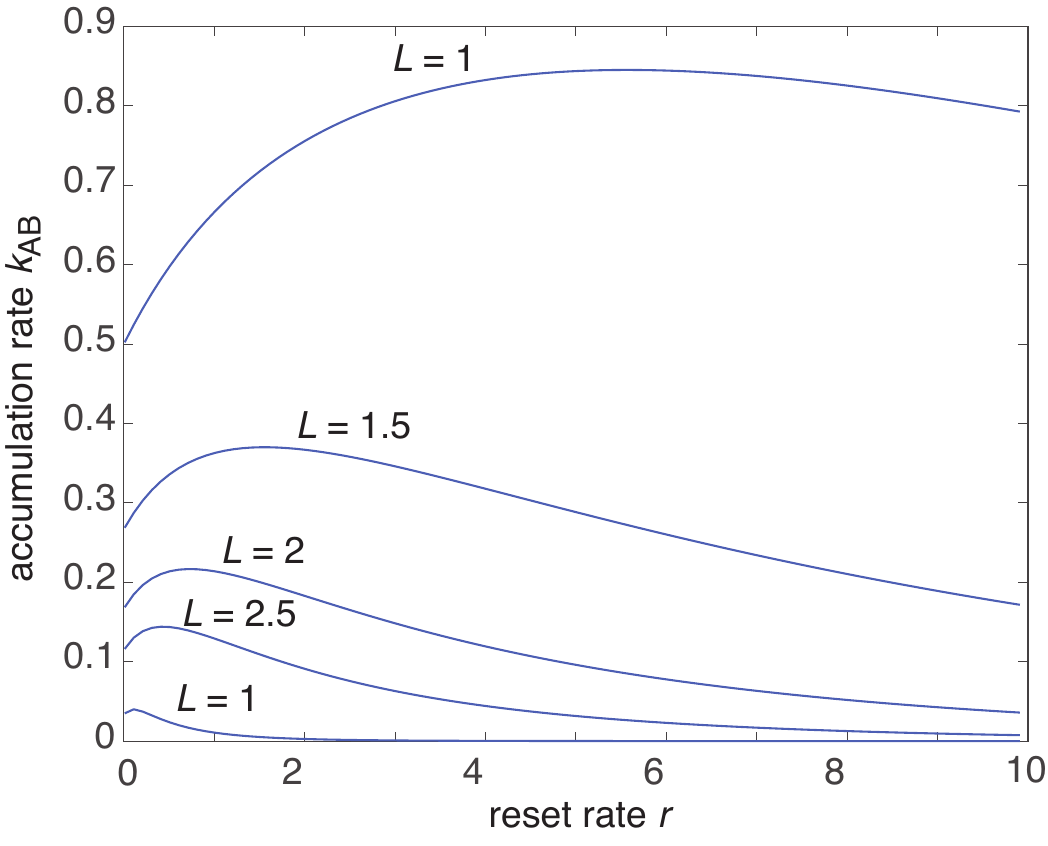}
\caption{Plot of target accumulation rate $k_{AB}$ as a function of the reset rate $r$ for various values of the domain size $L$ for $x_r=0.1$. Other parameters are $L=1$, $a=b=0.5$ and $D=1$.}
\label{fig10}
\end{figure}

Using the small-$r$ behavior of $\rho(x)$ and $q(x)$, we find from equation (\ref{final1D}) that 
\begin{equation}
\lim_{r\rightarrow 0}k_{AB}=\frac{1}{L+b+a}\frac{1}{L^2}.
\end{equation}
Moreover, in the limit $r\rightarrow \infty$, we have $k_{AB}\rightarrow 0$ since the particle resets so frequently that it never has a chance to collect resources and deliver them to the target. As with previous studies of search processes with resetting in the interval \cite{Christou15,Pal19}, we wish to determine whether or not there exists an optimal resetting rate at which $k_{AB}$ is maximized. This is explored in Fig. \ref{fig8}, where we plot $k_{AB}$ as a function of $r$ for $a=b=0.5$, $L=1$ and various values of $x_r$. (Our results are obtained by numerically integrating the various terms in equation (\ref{final}.) As expected, $k_{AB}\rightarrow 0.5$ as $r\rightarrow 0$. We find that for $0<x_r < 0.5$, there exists an optimal resetting rate, which is an increasing function of $x_r$. (Since the source and target domains have the same length, it follows that $k_{AB}$ is invariant under the mapping $x_r\rightarrow 1-x_r$. The latter no longer holds when $a\neq b$. This is illustrated in Fig. \ref{fig9} for the case $a>b$, which shows that if $x_r < 0.5$ then the accumulation rate for fixed $r$ increases under the mapping $x_r\rightarrow 1-x_r$. Such a result makes sense, since taking the reset position to be closer to the target means that the particle is less likely to waste time diffusing within the source domain. In Fig. \ref{fig10}, we plot $k_{AB}$ against $r$ for various lengths $L$ and fixed $x_r,a,b$. As expected, the accumulation rate is a a decreasing function of $L$. We also find that the optimal reset rate decreases with increasing $L$. Finally, note that in the limit $a,b\rightarrow \infty$ and $r=0$, the analysis breaks down since diffusion in an unbounded domain is non-ergodic.

%\setcounter{equation}{0}
%\renewcommand{\theequation}{A.\arabic{equation}}
%\section*{Appendix A: Evolution equation in Laplace space}

\section{Discussion} In this paper we developed a novel application of TPT to a dual-aspect search process, in which a diffusing particle first has to find a source domain $A$ in order to collect cargo, and then has to  find a distinct target domain $B$ where the cargo is delivered, see Fig. \ref{fig1}(b). For simplicity, we assumed that (i) cargo loading and unloading do not interrupt the ongoing search process, and (ii) the particle does not maintain a memory of the locations of $A$ and $B$. The mapping of the dual-aspect search process to TPT is based on the identification of the rate $k_{AB}$ at which the target accumulates resources with the time-averaged probability flux across a dividing surface between $A$ and $B$. The calculation of $k_{AB}$ then assumes that the underlying search process is ergodic. In this paper, we focused on the particular example of diffusion with stochastic resetting in the absence of an external potential. The calculation of $k_{AB}$ required taking into account of the fact that the stochastic process is not time-reversal invariant, and that transition paths can jump discontinuously across $S$ via resetting. In the case of diffusion in the interval, we established that the accumulation rate $k_{AB}$ is a nontrivial function of the reset rate $r$. In particular, there exists an optimal reset rate at which $k_{AB}$ is maximized.

There are a variety of possible extensions of the current work. First, one could consider higher-dimensional examples of the basic model developed in this paper. Second, one could explore what happens if assumption (i) or (ii) is relaxed. Third, there are a wide range of other stochastic processes that could be incorporated into the dual-aspect search process, most notably, active Brownian motion, run-and-tumble dynamics, and L\`evy flights. A third non-trivial extension is to consider multiple source and target domains. One of the difficulties is that it is no longer clear how to generalize the notion of a dividing surface. One possible approach would be to extend the sequential event analysis of Ref. \cite{Lorpaiboon22}.

\end{document}